\documentstyle[prl,aps] {revtex}                      %%%%%  new REVTeX 3.0

\def\Journal#1#2#3#4{{#1} {\bf #2}, #3 (#4)}
\def\NPA{{\em Nucl. Phys.} A}

\def\PLB{{\em Phys. Lett.}  B}
\def\PRL{\em Phys. Rev. Lett.}
\def\PRC{{\em Phys. Rev.} C}

\def\ZPA{{\em Z. Phys.} A}
\def\EPJA{{\em Eur. Phys. J.} A}
\def\PR{{\em Phys. Rev.}}

\def\JPG{\em J. Phys. G: Nucl. Part. Phys.}

\begin{document}
%\draft
\twocolumn
\title{A polarization relation and the measurement of the longitudinal response
       in pseudoscalar meson electroproduction off the nucleon}

\author{Hartmut Schmieden and Lothar Tiator}
\address{Institut f\"ur Kernphysik, Universit\"at Mainz, 55099 Mainz,
         Germany
        }

\date{\today}

\maketitle

\begin{abstract}
  For pseudoscalar meson electroproduction off the nucleon
  in parallel kinematics
  a relation between three polarization observables is derived. It is
  shown that, without Rosenbluth separation, a measurement of the
  longitudinal strength can be achieved through three different ways.
  They are discussed with preliminary MAMI data for the $p(\vec e,
  e'\vec p)\pi^0$ reaction in the energy range of the $\Delta(1232)$
  resonance.


keywords:  delta resonance, quadrupole excitation, recoil polarization

\end{abstract}

\section{Introduction}

In the past few years, elastic and quasielastic scattering experiments
yielded decisive results concerning the electromagnetic structure
of proton \cite{Jones99} and neutron
\cite{Eden94,Ostrick99,Herberg99,Passchier99,Rohe99}
by taking advantage of double polarization observables.
These provide also high sensitivity to the longitudinal
electromagnetic coupling of nucleon resonances \cite{HS98}.
Experimentally, these short lived  resonances can be tagged through
their decay into the nucleon-meson channel.
Pseudoscalar meson production is thus of particular interest for
resonance physics.
Due to the inherent problem of separation of non-resonant background,
unpolarized measurements are valuably complemented by polarization
experiments, 
which also profit from their insensitivity to major sources of
systematic uncertainties.

Furthermore, polarization observables can be utilized for a
separation of longitudinal and transverse response.
Thus, the experimental difficulties of the standard Rosenbluth-technique
\cite{Blomqvist96,Distler98} can be circumvented.
Three different methods for the extraction of the ratio of
longitudinal and transverse strength are discussed on the basis of
first, preliminary $p(\vec e, e'\vec p)\pi^0$  data
in the energy range of the $\Delta(1232)$  resonance \cite{HS99,Pospisch99}
from the Mainz Microtron MAMI.

\section{Recoil polarization in parallel kinematics}

\begin{figure}[t]
\centerline{\psfig{figure=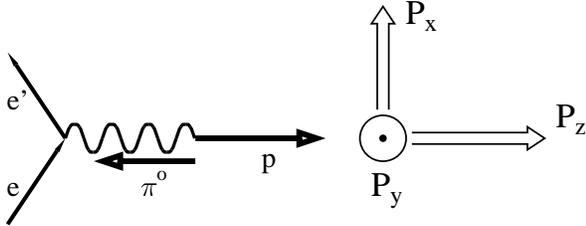,width=8.5cm}}
\caption{ Electroproduction of pseudoscalar mesons in parallel kinematics
          for the example of the $p(\vec e, e'p)\pi^0$  reaction.
          The components $P_{x,y,z}$  of the proton polarization are
          defined relative to the electron scattering plane.  }
\label{fig:kinematics}
\end{figure}

The present double-polarization experiments focus on the situation of
parallel kinematics, where the recoiling nucleon of the
$p(e,e'p)\pi^0$ reaction is detected along the direction of the
momentum transfer, $\vec q$. In this case the components of the recoil
nucleon polarization \footnote{Target polarization is equivalent to
  the measurement of recoil polarization when the cross-section
  asymmetry with regard to the reversal of beam helicity is
  considered.  } are given by:
\begin{eqnarray}
P_x &=& P_e \cdot c_- \cdot \frac{R_{LT'}^t}{R_T + \epsilon_L R_L}
\label{eq:P_x} \\
P_y &=& \qquad    c_+ \cdot \frac{R_{LT}^n }{R_T + \epsilon_L R_L}
\label{eq:P_y} \\
P_z &=& P_e \cdot c_0 \cdot \frac{R_{TT'}^l}{R_T + \epsilon_L R_L}
\label{eq:P_z}
\end{eqnarray}
The coordinate frame is defined relative to the electron scattering
plane as depicted in Fig.\ref{fig:kinematics}.
In contrast to ref.\cite{HS98} here we use the notation of Drechsel and
Tiator \cite{DT92}.
The structure functions, $R_K^i$, have to be taken at the pion cm-angle
of $\Theta_{\pi}^{cm} = 180^{\circ}$.
$P_e$  denotes the longitudinal electron polarization,
and the kinematical factors are
\begin{equation}
c_{\pm} = \sqrt{2 \epsilon_L (1 \pm \epsilon)} \quad \mbox{and}\quad
c_0     = \sqrt{1 - \epsilon^2},
\end{equation}
where $\epsilon=(1 + 2{\vec q}^{\,2}/Q^2
\tan^2\frac{1}{2}\vartheta_e)^{-1}$ and
$\epsilon_L=(Q^2/\omega_{cm}^2)\epsilon$ represent the degrees of
tranverse and longitudinal polarization of the virtual photon,
respectively. $Q^2$ is the negative squared four-momentum transfer,
$\vartheta_e$ the electron scattering angle and $\omega_{cm}$ the
energy transfer in the photon-nucleon cm frame.

From the components of recoil polarization
(Eqs.\ref{eq:P_x}-\ref{eq:P_z}) we define reduced polarizations
through normalization by the virtual photon polarization factors:
\begin{eqnarray}
\chi_x = \frac{1}{P_e c_-}\cdot P_x &=& \frac{R_{LT'}^t}{R_T+\epsilon_L R_L}
\label{eq:chi_x} \\
\chi_y = \frac{1}{c_+}\cdot P_y     &=& \frac{R_{LT}^n }{R_T+\epsilon_L R_L}
\label{eq:chi_y} \\
\chi_z = \frac{1}{P_e c_0}\cdot P_z &=& \frac{R_{TT'}^l}{R_T+\epsilon_L R_L}
\label{eq:chi_z}
\end{eqnarray}
The structure functions are conveniently expressed in terms of six
helicity amplitudes, $H_{1-6}$, or CGLN amplitudes, $F_{1-6}$
\cite{DT92,KDT95,CGLN57}. This yields:
\begin{eqnarray}
R_T       &=& \frac{1}{2} |H_4|^2 = | F_1 + F_2 |^2 
\label{eq:RT} \\
R_L       &=& |H_6|^2 =|F_5-F_6|^2\\
R_{LT'}^t &=& -\frac{1}{\sqrt{2}} \Re{e}\{H_6^* H_4\}  \nonumber \\
          &=& -\Re{e}\{(F_6^*-F_5^*) (F_1+F_2)\}  \\
R_{LT}^n  &=& \frac{1}{\sqrt{2}} \Im{m}\{ H_6^* H_4 \} \nonumber \\
          &=& \Im{m}\{(F_6^*-F_5^*) (F_1+F_2) \} \\
R_{TT'}^l &=& \frac{1}{2} |H_4|^2 = | F_1 + F_2 |^2 \label{eq:RTT'}
\end{eqnarray}
$R_{LT'}^t$  and $R_{LT}^n$  represent real and imaginary parts of the same
complex interference term.
The equality
\begin{equation}
R_T = R_{TT'}^l
\label{eq:equality}
\end{equation}
is a pecularity of parallel kinematics.
\input{psfig}

\section{Polarization relation and L/T separation}

From Eqs.\ref{eq:RT}-\ref{eq:RTT'} it can be easily seen that
\begin{equation}
\chi_x^2 + \chi_y^2 = \frac{\frac{1}{2}|H_4|^2 |H_6|^2}{(R_T+\epsilon_L R_L)^2}
                    = \frac{R_T \cdot R_L}{(R_T+\epsilon_L R_L)^2}
\label{eq:alpha}
\end{equation}
and
\begin{equation}
\chi_z^2 = \frac{R_T^2}{(R_T+\epsilon_L R_L)^2}.
\label{eq:1}
\end{equation}
Therefore, the ratio between longitudinal and transverse response is given by
\begin{equation}
\frac{R_L}{R_T} = \frac{\chi_x^2 + \chi_y^2}{\chi_z^2}.
\label{eq:2}
\end{equation}
The extraction of $R_L/R_T$  from Eq.\ref{eq:2} requires the measurement of
all three polarization components.
It can also be obtained from $\chi_z$  alone \cite{Kelly99}
by rewriting Eq.\ref{eq:1}:
\begin{equation}
\frac{R_L}{R_T} = \frac{1}{\epsilon_L}\left(\frac{1}{\chi_z}-1\right)
\label{eq:3}
\end{equation}
Combining Eqs.\ref{eq:3} and \ref{eq:2} we directly receive a model
independent relation between the three reduced polarizations:
\begin{equation}
\chi_x^2 + \chi_y^2 = \frac{1}{\epsilon_L} \chi_z(1-\chi_z)
\label{eq:sumrule}
\end{equation}
This equation relates the absolute value of the transverse polarization
with the longitudinal polarization.
It represents a constraint for any model.
For example, it is perfectly fullfilled by the Mainz Unitary Isobar Model
\cite{Drechsel99} for pion photo and electroproduction.

The relation can also serve as a consistency check for experimental data.
Up to now there is one prelimenary data set available
with all three components of proton polarization simultaneously measured
in the reaction
$p(\vec e, e'\vec p)\pi^0$ at the energy of the $\Delta$  resonance
\cite{HS99,Pospisch99}.
Within their present errors these data fullfill Eq.\ref{eq:sumrule}:
\begin{eqnarray}
\chi_x^2 + \chi_y^2 &=& 0.0348 \pm 0.0045_{\mbox{\tiny stat}}
                               \pm 0.0031_{\mbox{\tiny syst}} = \nonumber\\
\frac{1}{\epsilon_L} \chi_z(1-\chi_z)
                    &=& 0.0519 \pm 0.022_{\mbox{\tiny stat}}
                               \pm 0.007_{\mbox{\tiny syst}}
\label{eq:sumrule_exp}
\end{eqnarray}
We note, however, that the rhs of Eq.\ref{eq:sumrule_exp} does not produce
a very strong constraint.
This is due to the unfavorable error propagation when the small deviation
of $\chi_z$  from unity is measured.

The determination of $R_L/R_T$  from Eq.\ref{eq:3} suffers by the same
reason.
We obtain:
\begin{equation}
\frac{R_L}{R_T} = 0.066 ^{+0.038}_{-0.034_{\mbox{\tiny stat}}}
                         \pm 0.011_{\mbox{\tiny syst}}
\end{equation}
A relative statistical error of 6.2\,\% in $\chi_z$
is amplified to 57.6\,\% in $R_L/R_T$.
This is a consequence of the weak influence of $R_L$  on the longitudinal
polarization component.
Therefore, in the $\Delta(1232)$  region
$P_z$  should rather be used as an experimental consistency check,
because it is almost entirely determined by the polarization of the
electron beam and electron kinematics \cite{HS98}:
\begin{equation}
P_z \simeq P_e \cdot c_0 %\sqrt{1-\epsilon^2}
\end{equation}

A better way to determine the longitudinal strength is through Eq.\ref{eq:2},
which yields
\begin{equation}
\frac{R_L}{R_T} = 0.044
                  \pm 0.006_{\mbox{\tiny stat}} \pm 0.004_{\mbox{\tiny syst}}.
\end{equation}
However, all three polarization components need to be measured simultaneously.
A procedure which requires only the two transverse reduced polarizations
starts from Eq.\ref{eq:alpha}
as a quadratic equation for $R_L/R_T$.
The solutions are
\begin{equation}
\frac{R_L}{R_T} = \left[ \alpha_L \pm \sqrt{\alpha_L^2-\epsilon_L^2} \right]^{-1},
\label{eq:LT_solution}
\end{equation}
where
\begin{equation}
\alpha_L = \frac{1}{2(\chi_x^2+\chi_y^2)} - \epsilon_L.
\end{equation}
To the extent that $R_L/R_T \ll 1/\epsilon_L$, Eq.\ref{eq:LT_solution} can be
simplified to
\begin{equation}
\frac{R_L}{R_T} = \frac{1}{2 \alpha_L}.
\label{eq:LT_short}
\end{equation}
The second solution yields $R_L > R_T$  and is thus obviously unphysical.
With Eq.\ref{eq:LT_short} the preliminary MAMI data \cite{HS99}
yield the result
\begin{equation}
\frac{R_L}{R_T} = 0.040
                  \pm 0.006_{\mbox{\tiny stat}} \pm 0.004_{\mbox{\tiny syst}}.
\end{equation}
For the latter two approaches it is possible to practically maintain the
experimental errors of the reduced polarizations in the extracted ratio.
In contrast to the extraction through Eq.\ref{eq:3} there is no error
amplification, but the measurement of two or three polarization
components is required.
Experimentally, the false systematic asymmetries of the recoil polarimeter
need to be under control. 
While they can be eliminated in the electron-helicity dependent components 
$P_x$  and $P_z$, they influence the extraction of $P_y$.

\section{Summary and conclusions}

For the case of electroproduction of pseudoscalar mesons off the nucleon
in parallel kinematics
with longitudinally polarized beam and with either measurement
of recoil nucleon polarization or target polarization a set of three
reduced polarizations has been defined.
There exists a model independent relation between the quadratic sum
of the two transverse reduced polarizations and the longitudinal one.
It puts a constraint on both phenomenological models and experimental data.

The polarization observables offer the possibility to measure the ratio of
longitudinal to transverse strength without the need of a
Rosenbluth-separation.
Three different ways have been discussed which require the measurement of
one, two or three of the reduced polarizations.
While the L/T-ratio from $\chi_z$  alone suffers from a very unfavourable
error propagation, in particular
the quadratic sum $\chi_x^2+\chi_y^2$  is  well suited for a measurement of
the longitudinal response.

\acknowledgments

The authors are indebted to Thomas Pospischil for sharing preliminary
experimental results of his doctoral thesis prior to publication.
This work was supported by the Deutsche Forschungsgemeinschaft (SFB 443).


\begin{thebibliography}{99}
\bibitem{Jones99} M.K. Jones et al., nucl-ex/9910005,
                  subm. to {\it Phys. Rev. Lett.}
\bibitem{Eden94} T. Eden et al., \Journal{\PRC}{50}{R1749}{1994}
\bibitem{Ostrick99} M. Ostrick et al., \Journal{\PRL}{83}{276}{1999}
\bibitem{Herberg99} C. Herberg et al., \Journal{\EPJA}{5}{131}{1999}
\bibitem{Passchier99} I. Passchier et al., \Journal{\PRL}{82}{4988}{1999}
\bibitem{Rohe99}    D. Rohe et al., \Journal{\PRL}{83}{4257}{1999}
\bibitem{HS98}      H. Schmieden, \Journal{\EPJA}{1}{427}{1998}
\bibitem{Blomqvist96} K.I. Blomqvist et al.,
                      \Journal{\ZPA}{353}{415}{1996} and
                      A. Liesenfeld et al.,
                      \Journal{\PLB}{468}{20}{1999}
\bibitem{Distler98}   M.O. Distler et al., \Journal{\PRL}{80}{2294}{1998}
\bibitem{HS99}      H. Schmieden, Proceedings of PANIC'99 and
                    nucl-ex/9909006
\bibitem{Pospisch99}    Th. Pospischil, doctoral thesis, Mainz, in preparation
\bibitem{DT92}      D. Drechsel and L. Tiator,
                    \Journal{\JPG}{18}{449}{1992}
\bibitem{KDT95}    G. Kn\"{o}chlein, D. Drechsel and L. Tiator,
                   \Journal{\ZPA}{352}{327}{1995}
\bibitem{CGLN57}    G.F. Chew et al., \Journal{\PR}{106}{1345}{1957}
\bibitem{Kelly99}   J.J. Kelly, \Journal{\PRC}{60}{054611}{1999}
\bibitem{Drechsel99}    D. Drechsel, O. Hanstein, S.S. Kamalov and L. Tiator,
                        \Journal{\NPA}{645}{145}{1999}

\end{thebibliography}
\end{document}